\documentclass[twoside,12pt]{article}
\usepackage{epsfig}
\def\Journal#1#2#3#4{{#1} {#2} (#4) #3 }

\def\NPA{{\em Nucl. Phys.} A}

\def\PRO{{\em Prog. Theor. Phys.}}

\def\PLB{{\em Phys. Lett.} B}

\def\PRL{\em Phys. Rev. Lett.}

\def\PREP{\em Phys. Rep.}

\def\PRD{{\em Phys. Rev.} D}
\def\PRC{{\em Phys. Rev.} C}

\def\ZPA{{\em Z. Phys.} A}
\def\JPG{{\em J. Phys.} G}
\def\EPJA{{\em Eur. J. Phys.} A}
\def\ANNP{\em Ann. Phys. (N.Y.)}

\def\ADN{\em Adv. Nucl. Phys.}
\def\HIP{\em Heavy Ion Phys.}

\begin{document}
\newcommand{\p}{\partial}
\newcommand{\D}{\Delta}
\newcommand{\ls}{\left(}
\newcommand{\rs}{\right)}
\newcommand{\beq}{\begin{equation}}
\newcommand{\eeq}{\end{equation}}
\newcommand{\beqa}{\begin{eqnarray}}
\newcommand{\eeqa}{\end{eqnarray}}
\newcommand{\bdm}{\begin{displaymath}}
\newcommand{\edm}{\end{displaymath}}
\newcommand{\fps}{f_{\pi}^2 }
\newcommand{\mks}{m_{{\mathrm K}}^2 }
\newcommand{\ms}{m_{{\mathrm K}}^{*} }
\newcommand{\mk}{m_{{\mathrm K}} }
\newcommand{\msq}{m_{{\mathrm K}}^{*2} }
\newcommand{\rhos}{\rho_{\mathrm s} }
\newcommand{\rhob}{\rho_{\mathrm B} }
\title{\vspace{1cm} Dilepton production in elementary
and in heavy ion reactions} 
\author{Christian Fuchs and Amand Faessler\\
Institut f\"ur Theoretische Physik\\ 
Universit\"at T\"ubingen, D-72076 T\"ubingen, Germany }
\date{}
\maketitle
\begin{abstract}
We present a unified description of the vector 
meson and dilepton production in elementary 
and in heavy ion reactions. The production of vector mesons 
($\rho,\omega,\phi$) 
is described via the excitation of nucleon resonances ($R$). 
The theoretical framework is an extended vector meson dominance 
model (eVMD) for resonance decays $R\longmapsto NV$ with 
arbitrary spin which is covariant and kinematically complete. The eVMD 
includes thereby excited vector meson states  in the 
transition form factors. The model 
has successfully been applied to $\omega$ and $\phi$ 
production in $p+p$ reactions. 
The same model is used to describe the 
dilepton production in elementary reactions where corresponding 
data are well reproduced. However, when the model is applied to 
heavy ion reactions in the BEVALAC/SIS energy range  
the experimental dilepton spectra measured by the DLS Collaboration 
are significantly underestimated at small invariant masses. 
In view of this fact we discuss further medium effects:  
One is a substantial collisional 
broadening of the $\rho$ and in particular of the $\omega$ meson in 
the vicinity of the $\rho/\omega$-peak. The second medium effect 
is the destruction of 
quantum interference in a dense medium. A decoherent 
dilepton emission through vector mesons decays enhances the corresponding 
low mass dilepton yield in heavy ion reactions and improves the 
agreement with existing data. 
\end{abstract}
\section{Introduction}
One of the important questions which theorists face at present 
is the dependence of hadron properties on medium effects. 
Medium effects manifest themselves in the modification of 
widths and masses of resonances produced in nuclear collisions.
The magnitude of such changes depends thereby on the density and 
the temperature of the medium. 
E.g., the proposed Brown-Rho scaling \cite{BR} is equivalent to a reduction 
of the vector meson masses in the nuclear medium. The same conclusion 
is obtained from QCD sum rules \cite{QCDSR}
and within effective hadronic models \cite{BGP}. The dispersion 
analysis of forward scattering amplitudes 
\cite{KKW,EIoffe,EBEK,KS} showed that vector meson mass shifts are in 
general small and positive, whereas at low momenta they can change 
the sign which is in qualitative agreement with the 
Brown-Rho scaling and the results from QCD sum rules. However, 
the question of in-medium masses must finally 
be settled experimentally. 

Dilepton spectra from heavy-ion collisions are considered 
as a suitable tool for this
purpose. The CERES \cite{ceres} and HELIOS \cite{HELIOS} 
Collaborations measured dilepton spectra at CERN and found
a significant enhancement of the low-energy dilepton yield below the 
$\rho $ and $\omega $ peaks \cite{ceres} in heavy reaction systems 
($Pb+Au$) compared to light systems ($S+W$) and 
proton induced reactions ($p+Be$). Theoretically, this enhancement 
can be explained within a hadronic picture by the assumption of 
a dropping $\rho $ mass \cite{drop} or by the inclusion 
of in-medium spectral functions for the vector mesons 
\cite{rapp,BCRW98}. In both cases the enhanced low energetic 
dilepton yield is not simply caused by a shift of the 
$\rho $ and $\omega $ peaks in the nuclear medium but it originates 
to most extent from an enhanced contribution of the $\pi^+ \pi^-$ annihilation 
channel which, assuming vector dominance, runs over an intermediate 
$\rho$ meson. An alternative scenario could be the formation 
of a quark-gluon plasma which leads to additional 
($pQCD$) contributions to the dilepton spectrum 
\cite{rapp,weise00}.

A similar situation occurs at a completely different energy scale,
 namely around 1 A.GeV incident energies where the low mass region
of dilepton spectra are underestimated by present transport
calculations compared to $pp$ and $pd$ reactions. The corresponding 
data were obtained by the DLS Collaboration at the BEVALAC \cite{DLS}. 
However, in contrast to ultra-relativistic reactions (SPS) 
the situation does not improve when full spectral functions
and/or a dropping mass of the vector mesons are taken
into account \cite{ernst,BK,BCRW98}. This fact is known as 
the DLS {\it puzzle}. The reason lies in the fact 
that both, possible $pQCD$ contributions as well as a sufficient amount of 
$\pi^+ \pi^-$ annihilation processes are absent at intermediate energies. 
Also a dropping $\eta$ mass 
can be excluded as a possible explanation of the DLS puzzle since it 
would contradict $m_T$ scaling \cite{BCRW98}. Furthermore, chiral 
perturbation theory predicts only very small modifications of the 
in-medium $\eta$ mass \cite{oset02}. 
Thus one has to search for other sources which could explain the low mass 
dilepton excess seen in heavy ion reactions. 
Dilepton spectra were also measured at KEK in $p + A$ reactions at a beam 
energy of 12 GeV \cite{KEK}. Also here an excess of 
dileptons compared to the known sources was 
observed below the $\rho$-meson peak and interpreted as a change of the 
 vector meson spectral functions. These data were recently analyzed in 
Ref. \cite{Elena},  again without success to explain the experimental 
spectrum within a  dropping mass scenario and/or by a 
significant collision broadening
of the vector mesons. Since the vector meson peaks are not 
resolved experimentally \cite{DLS}, the problem to extract in-medium 
masses directly from experimental data remains extremely difficult. 

For all these studies a precise and rather complete knowledge of the relative
weights for existing decay channels is indispensable in order to
draw reliable conclusions from dilepton spectra. 
In \cite{krivo00} a systematic study of meson decay channels
was performed, including channels which have been neglected so far, 
such as e.g. four-body decays $\rho^0\rightarrow\pi^0\pi^0e^+e^-$. 
However, as has been shown in \cite{resdec} in $pp$ reactions 
the contributions of these more exotic channels are not large enough 
to enhance the low mass dilepton yield at incident energies around 
1 AGeV. Here the low mass dilepton spectrum is dominated by the $\eta$ 
and the contributions from the decay of baryonic resonances 
\cite{ernst,resdec,BCEM}. 

The importance of the resonance contribution to the dilepton 
yield in elementary and heavy ion reactions has been stressed in several works 
\cite{resdec,koch96,pirner,post01,titov,zetenyi,BCM,resgiessen2,lutz03,mosel03,krivo01,krivo02}.  
In \cite{krivo02} we calculated in a fully relativistic 
treatment of the dilepton decays $R\rightarrow N\ e^+\ e^-$ of nucleon 
resonances with masses below 2 GeV. Kinematically complete 
phenomenological expressions for the dilepton decays of 
resonances with arbitrary spin and parity,
parameterized in terms of the magnetic, electric, and Coulomb transition form
factors and numerical estimates for the dilepton spectra and branching
ratios of the nucleon resonances were given. In \cite{resdec} this approach 
was applied to the dilepton production in $pp$ reactions at BEVALAC 
energies. The relevant elementary hadronic reactions 
were systematically discussed. It is demonstrated that 
the resonance model provides an accurate description of 
exclusive vector meson production in nucleon-nucleon 
collisions $NN\rightarrow NN\rho(\omega)$ as well as in pion scattering 
$\pi N\rightarrow N\rho(\omega)$. 
The resonance model allows further to determine the isotopic channels of the 
$N N\rightarrow N N\rho(\omega)$ cross section where no data 
are available. As discussed in \cite{omega02}, a peculiar role plays thereby the 
$N^*(1535)$ resonance which, fitting available photo-production data, has 
a strong coupling to the $N\omega$ channel. Close to threshold this can lead to 
strong  off-shell contributions to the  $\omega$ production cross section 
\cite{omega02} which are also reflected in the dilepton yields.

The reaction dynamics of heavy ion collisions is described 
within the QMD transport model \cite{shekther03} which has 
been extended, i.e. the complete set of baryonic 
resonances ($\Delta$ and $N^*$) with masses below 2 GeV has 
been included in the T\"ubingen transport code. 
One purpose of the present investigations is to extract 
information on the in-medium $\rho$- and $\omega$-meson 
widths directly form the BEVALAC data \cite{DLS}. 
The dilepton spectra, distinct from the vector meson masses,
are very sensitive to the vector meson in-medium widths, 
especially the $\omega$-meson. 
The collision broadening is a universal mechanism to increase particle
widths in the medium. E.g., data on the total
photo-absorption cross section on heavy nuclei \cite{fras} provide 
evidence for a broadening of nucleon resonances in a nuclear medium \cite
{Kondratyuk:1994ah}. The same effect should
be reflected in a broadening of the vector mesons in dense matter. 
Since the DLS data show no peak structures which can be attributed 
to the vector meson masses, the problem to extract information 
on possible mass shifts is not yet settled. However, the data 
allow to estimate the order of magnitude of the collision broadening 
of the vector mesons in heavy ion collisions. 

Another  question is the role of 
 quantum interference effects. Semi-classical transport models 
like QMD do not keep track of relative phases between amplitudes but 
assume generally that decoherent probabilities can be propagated. On the other 
hand, it has been stressed in several works \cite{zetenyi,lutz03} 
that, e.g., the interference of the isovector-isoscalar 
channels, i.e. the so-called $\rho-\omega$ mixing can significantly alter 
the corresponding dilepton spectra. The $\rho-\omega$ mixing was 
mainly discussed for the dilepton production in $\pi N$ reactions. 
Due to the inclusion of excited mesonic states in the resonance 
decays such interference occurs in our treatment already separately inside each 
isotopic channel. It is natural to 
assume that the interference pattern of the mesonic states will be 
influenced by the presence of surrounding particles. 
In \cite{shekther03} we discussed decoherence effects which 
can arise when vector mesons propagate through a hot and dense medium
 and proposed a simple scheme to model this type of decoherence
 phenomenon. This discussion is 
quite general and can be applied, e.g. to the $\rho-\omega$ mixing  
as a special case. It is assumed that before the first 
collision with a nucleon or a pion 
the vector mesons radiate $e^+e^-$ pairs coherently and decoherently afterwards,
since the interactions with a heat bath result in macroscopically 
different final states. As a consequence of charge conservation the 
coherence must be restored in the soft-dilepton limit. The 
present model fulfills this boundary condition. 
The quark counting rules require a destructive interference
between the vector mesons entering into the electromagnetic 
transition form factors of the nucleon resonances. 
Hence, a break up of the coherence results in an increase of 
the dilepton yield below the $\rho$-meson peak. This is 
just the effect observed in the BEVALAC data.
Such a quantum decoherence can at least partially 
resolve the DLS puzzle in heavy ion reactions.

\section{Resonance model and extended VMD}

Usually, the description of the decays of baryonic resonances 
$R\rightarrow N\ e^+\ e^-$ is based on the VMD model in its
monopole form, i.e. with only one virtual vector meson ($V = \rho,~\omega$).
As the result, the model provides a consistent description of both, 
radiative $R\rightarrow N\gamma$ 
and mesonic $R\rightarrow NV$ decays. However, a normalization 
to the radiative branchings strongly
underestimates the mesonic ones \cite{resdec,post01,pirner}. 
Possible ways to circumvent this 
inconsistency were proposed in \cite{pirner,post01}. In \cite{pirner} 
a version of the VMD model with vanishing $\rho\gamma$ coupling in the 
limit of real photons ($M^2=0$) was used which allows to fit 
radiative and mesonic decays independently, in \cite{post01} an 
additional direct coupling of the resonances to photons was introduced.  
\begin{figure}[h]
\begin{minipage}[h]{105mm}
\unitlength1cm
\begin{picture}(10.,4.0)
\put(0.5,0){\makebox{\epsfig{file=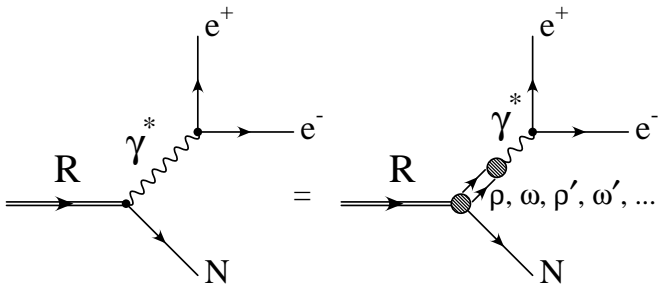,width=10.0cm}}}
\end{picture}
\end{minipage}
\vspace*{-1.5cm}
\hspace*{1.0cm}
\begin{minipage}[h]{60mm}
\caption{Decay of nuclear resonances to dileptons in the extended VMD 
model. The $RN\gamma$ transition form factors contain contributions from 
ground state and excited $\rho$ and $\omega$ mesons.
}
\end{minipage}
\label{graph1_fig}
\vspace*{1.5cm}
\end{figure}

However, apart from that the standard VMD predicts a
$1/t$ asymptotic behavior for the transition form factors. At the 
same time the quark counting rules require a stronger suppression at 
high $t$. A similar problem arises with the $\omega$ Dalitz decay. 
The  $\omega \pi \gamma $ transition form factor shows an 
asymptotic $\sim 1/t^{2}$ behavior \cite{VZ}. It has been 
measured in the time-like region \cite{LGL} and the
data show deviations from the naive one-pole approximation. 
In \cite{krivo00} it was shown that the inclusion of 
higher vector meson resonances in the VMD can resolve this problem 
and provides the correct asymptotics. In \cite{krivo02} the
extended VMD (eVMD) model was used to describe the decay of baryonic 
resonances and in particular to solve the inconsistency between $RNV$ and
$RN\gamma$ decay rates. In the  eVMD model one assumes that 
radial excitations $\rho (1250),~ \rho (1450),\dots$ can 
interfere with the ground state $\rho$-meson in radiative processes. Already in the 
case of the nucleon form factors the standard VMD is not sufficient 
and radially excited vector mesons 
$\rho ^{\prime },$ $\rho ^{\prime \prime }$ $...$ etc. should be
added in order to provide a dipole behavior of the Sachs
form factors and to describe the experimental data.
In view of these facts the present extension of the VMD model  is 
more general than the approach pursued in \cite{pirner} since it 
allows not only to describe consistently resonance decays but also 
other observables like the $\omega$ Dalitz decay or the nucleon 
form factor. Here we only briefly sketch the basic 
ideas of the extended vector 
meson dominance (eVMD) model. In Fig. 1 the resonance 
decays are schematically displayed for the extended VMD model 
with excited mesons as intermediate states. The interference 
between the different meson families 
plays a crucial role for the behavior of the form factors. 
Details of the relativistic calculation 
of the  magnetic, electric, and Coulomb transition form
factors and the branching ratios of the nucleon resonances 
can be found in \cite{krivo02}.

The vector meson production cross section in elementary
nucleon-nucleon reactions is now given by 
\begin{equation}
\frac{d\sigma (s,M)^{NN\rightarrow NNV}}{dM^{2}}
=\sum_{R}\int_{(m_{N}+M)^{2}}^{(\sqrt{s}-m_{N})^{2}}d\mu ^{2}\frac{ 
d\sigma (s,\mu )^{NN\rightarrow NR}}{d\mu ^{2}}
\frac{dB(\mu,M)^{R\rightarrow VN}}{d M^{2}}~~.
\label{sigNNV}
\end{equation}
The cross sections for the resonance production are given by 
\begin{equation}
d\sigma (s,\mu )^{NN\rightarrow NR} = 
\frac{|{\cal M}_R|^2 ~ p_f}{16 p_i s\pi}~dW_R(\mu)
\label{sigNR}
\end{equation}
with the final c.m. momentum 
\begin{eqnarray}
p_f = p^*(\sqrt{s},\mu,m_{N})
= \frac{\sqrt{(s-(\mu+m_N)^2)(s-(\mu-m_N)^2)}}{2\sqrt{s}}
\label{cr4}
\end{eqnarray}
and the initial c.m. momentum $p_i$. The mass distributions 
$dW_R(\mu)$ of the resonances are usual Breit-Wigner distributions 
\begin{equation}
dW_R(\mu) = 
\frac{1}{\pi} \frac{\mu \Gamma^R (\mu) d\mu^2 }
{(\mu^2 - m_{R}^2)^2 +(\mu\Gamma_{\rm tot}^R(\mu))^2}
\label{BW}
\end{equation}
where $\mu$ and $m_R$ are the running and pole masses, respectively, 
and $\Gamma(\mu)$ is the mass dependent resonance width. The 
matrix elements ${\cal M}_R$ are taken from \cite{Teis,Bass} where 
they have been adjusted to one and two-pion production data. 
For the description of the $\rho$ and $\omega$ production in $NN$ and 
$\pi N$ reactions we consider the same set of 
resonances which has been used in refs. \cite{resdec,omega02}. It 
includes only the well established ($4*$) resonances listed by 
the PDG \cite{pdg} and is smaller than the complete set of 
resonances included in the QMD model. This set of resonances is, 
however, sufficient to describe the $NN$ and $\pi N$ vector meson
production data. 

\begin{figure}[h]
\begin{minipage}[h]{95mm}
\unitlength1cm
\begin{picture}(9.,9.0)
\put(0.5,0){\makebox{\epsfig{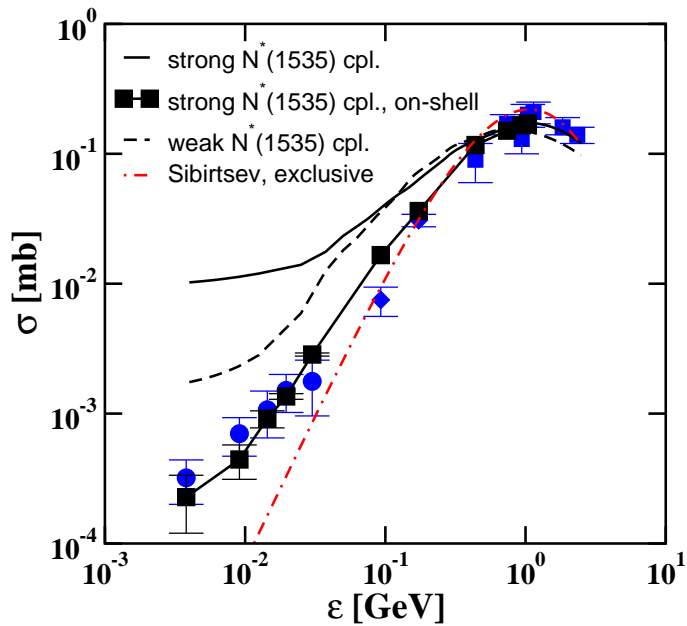}}}
\end{picture}
\end{minipage}
\vspace*{-1.5cm}
\hspace*{1.0cm}
\begin{minipage}[h]{70mm}
\caption{Exclusive $pp\rightarrow pp\omega$ cross section 
obtained in the resonance model as a function of the excess energy 
$\epsilon$. The solid curve shows the full cross 
section including off-shell 
contributions while the squares show the experimentally 
detectable on-shell part of the cross section. The dashed 
curves show the corresponding cross section obtained with weak 
$N^*(1535)N\omega$ coupling. The dot-dashed curve is a 
parameterization of the exclusive cross section from 
\protect\cite{sibirtsev97}. 
Data are taken from \protect\cite{hibou99,cosy01} and 
\protect\cite{disto01,flaminio}.
}
\end{minipage}
\vspace*{1.5cm}
\label{sigom_fig}
\end{figure}

Fig.\ref{sigom_fig} shows the 
$\omega$ production in elementary $NN$ reactions. The different cross 
sections are shown as functions of the excess energy $\epsilon$. 
As discussed in \cite{omega02}, the 
resonance model (with a large $N^*(1535)N\omega$ coupling) leads to 
very accurate description of the measured on-shell cross section. 
It has, however, a very strong off-shell component which fully 
contributes to the dilepton production. The weak coupling scenario, on 
the other side, has only small off-shell component but the 
reproduction of the data is relatively poor in the low energy regime. 
The parameterization of 
\cite{sibirtsev97} which has been used in \cite{BCRW98,BCEM} is also 
shown for comparison. 

In ref. \cite{phi03} the resonances model has been applied to the 
the production of $\phi$ mesons which, according to the 
``OZI rule''\cite{OZI}, should be strongly suppressed compared to that of
$\omega$ mesons. According to the OZI rule $\phi$ mesons can only be produced 
due to a small admixture of non-strange light quarks in their wave
function. The corresponding mixing angle $\theta_{mix}$
is equal to $\theta_{mix} \approx 3.7^o$. 
Experimentally, the ratio $R_{\phi/\omega}$ is in $pp\rightarrow
pp\phi(\omega)$  reactions, however, known to be one order of 
magnitude larger than the naive expectation. In ref. \cite{phi03} it 
is demonstrated that the experimental data are well 
reproduced by the present resonance model 
without introducing additional model parameters. 
\section{Dilepton production} 
\subsection{Elementary reactions}
Before turning to heavy ion collisions we will consider the dilepton 
production in elementary reactions. Dilepton spectra  in 
proton-proton and proton deuteron reactions have been measured by 
the DLS Collaboration in the energy range from $T=1\div 5$ GeV \cite{DLS2}. 
The application of the present model to the dilepton production 
in $pp$ reactions has in detail been discussed in \cite{resdec}. 
We show the corresponding results and the 
comparison to the DLS data \cite{DLS2} in Fig. 3. 
The agreement with the available data is generally reasonable, 
i.e. of similar quality as obtained in previous calculations by 
Ernst et al. \cite{ernst} and Bratkovskaja et al. \cite{BCM}. 
\begin{figure}[h]
\begin{minipage}[h]{105mm}
\unitlength1cm
\begin{picture}(10.,15.0)
\put(0.5,0){\makebox{\epsfig{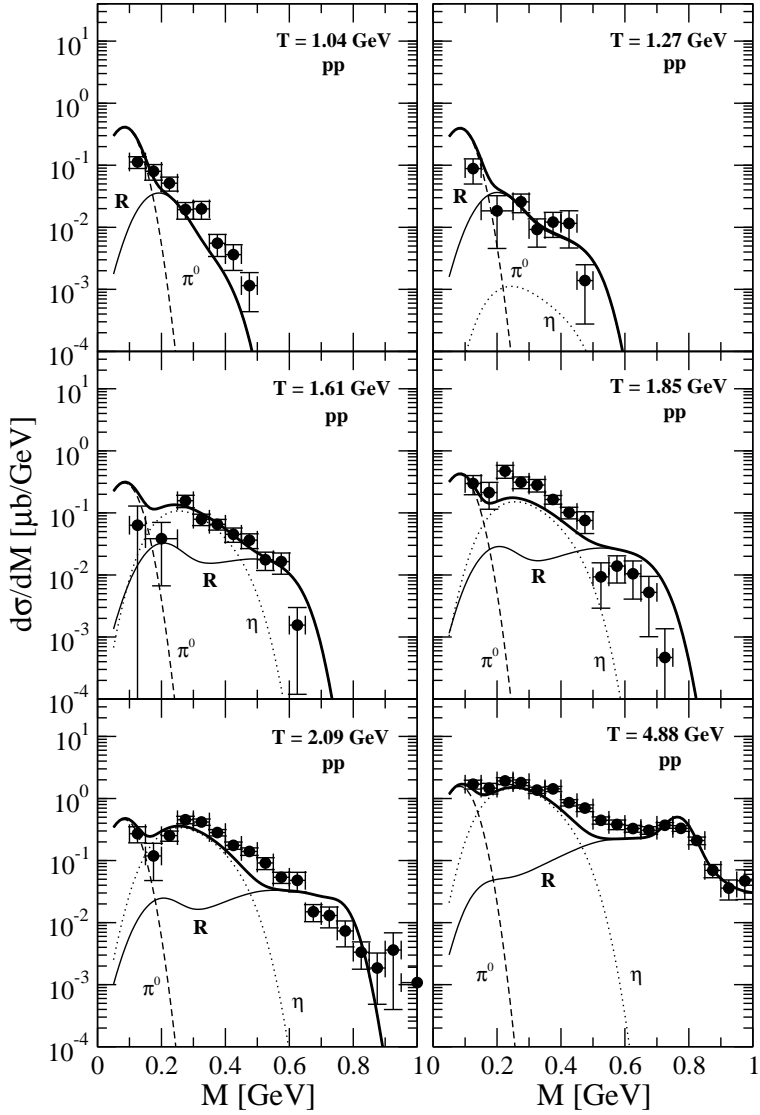}}}
\end{picture}
\end{minipage}
\vspace*{-1.5cm}
\hspace*{1.0cm}
\begin{minipage}[h]{50mm}
\caption{The differential $pp\rightarrow e^+e- X$ cross sections 
at various proton kinetic energies are compared to the DLS data 
\protect\cite{DLS2}. 
}
\end{minipage}
\label{DLS_fig1}
\vspace*{1.5cm}
\end{figure}
It should be noted that the dilepton yields in $pp$ reactions were 
obtained with the strong $N^*(1535)-N\omega$ decay mode. As in 
detail discussed in \cite{omega02} the 
strong coupling mode is the result of the eVMD fit to the 
available photo- and meson-production data \cite{krivo02}. 
It leads to sizable contributions from off-shell $\omega$ production around threshold 
energies which are, however, experimentally not accessible in 
$pp\rightarrow pp\omega$ measurements. On the other side, these off-shell 
$\omega$'s fully contribute to the dilepton yield. The off-shell 
contributions lead generally to an enhancement of the dilepton 
yield in the mass region below the $\omega$ peak, in particular 
at incident energies where the $\omega$ is dominantly produced 
subthreshold. 

The situation becomes more complicated when proton-deuteron 
reactions are considered. Compared to the $pp$ case one has here 
two important modifications: First the Fermi motion of the 
proton and neutron constituents inside the 
deuteron and secondly, the isotopic relations between
the $p p$ and $p n$ contributions to the dilepton production.
Only few isotopic relations for the meson production are 
experimentally fixed. Most isospin relations have to be derived 
from model assumptions. As shown in \cite{shekther03} 
the present model reproduces the dilepton production 
in $pd$ collisions at $T=1.61\div 4.88$ GeV rather reasonable. 
At the two lowest energies $T =1.04;~1.27$ GeV  we underestimate
the $pd$ data (probably due to an underestimation of the $\eta$ 
contribution). At these energies an underestimation which is, however, 
less pronounced, was also observed in \cite{ernst}. It should 
be noted that for the $pp$ reactions the present results and those of 
 \cite{ernst,BCM} coincide more or less. In all cases the 
theoretical calculations reproduce the corresponding DLS data 
reasonably well. Hence the dilepton production on the deuteron 
turns out to be rather involved at subthreshold energies due to 
strong ISI/FSI effects. The $pd$  
system is therefore only of limited use to check isospin relations 
of the applied models. Another important result is the fact that 
the scenario of large off-shell $\omega$ contributions from the 
$N^*(1535)-N\omega$ decay is consistent with the 
available $pp$ and $pd$  dilepton data.

\subsection{Heavy ion reactions}
With this input QMD transport calculations for $C+C$ 
and $Ca+Ca$ reactions at 1.04 AGeV have been performed \cite{shekther03}. 
First we discuss the results obtained without any additional medium 
effects concerning the dilepton production. For the nuclear mean field 
a soft momentum dependent Skyrme force (K=200 MeV) is used 
which provides also a good description of the subthreshold $K^+$ 
production in the considered energy range \cite{fuchs01}. 
The reactions are treated as minimal bias collisions with maximal impact 
parameters $b_{\rm max}=5(8)~{\rm fm}$ for $C+C (Ca+Ca)$. 

In Fig. 4 the results are compared to the DLS 
data. The acceptance filter functions provided by the 
DLS Collaboration are applied and the results are smeared over the 
experimental resolution of $\Delta M = 35$ MeV.  The calculations are 
performed within the two scenarios discussed in Sec. II, namely a 
strong  $N^*(1535)-N\omega $ coupling as implied by the original fit to 
the available photo-production data \cite{omega02} and a weaker coupling 
which can be enforced by a different choice of input parameters. In the 
first case strong off-shell $\omega$ contributions appear which are also 
visible in the dilepton spectrum at invariant masses below the 
$\omega$ peak. In the mass region between $0.4\div 0.8$ GeV the two 
scenarios yield significantly different results. The rest of the 
spectrum is practically identical except from the height of the $\omega$ 
peak itself. As discussed in connection with the elementary cross 
sections the $\omega$ contribution from the $N^*(1535)$ is suppressed 
at the  $\omega$ pole in the strong coupling scenario and thus the 
total  $\omega$ peak is slightly lower. The comparison of the transport calculations 
with the DLS data is here not completely conclusive: The lighter $C+C$ 
system would favor the weak  $N^*(1535)-N\omega $ coupling scenario 
whereas the $Ca+Ca$ reactions are better described by the strong coupling. 
\begin{figure}[h]
\begin{minipage}[h]{125mm}
\unitlength1cm
\begin{picture}(12.,7.0)
\put(0.5,0){\makebox{\epsfig{file=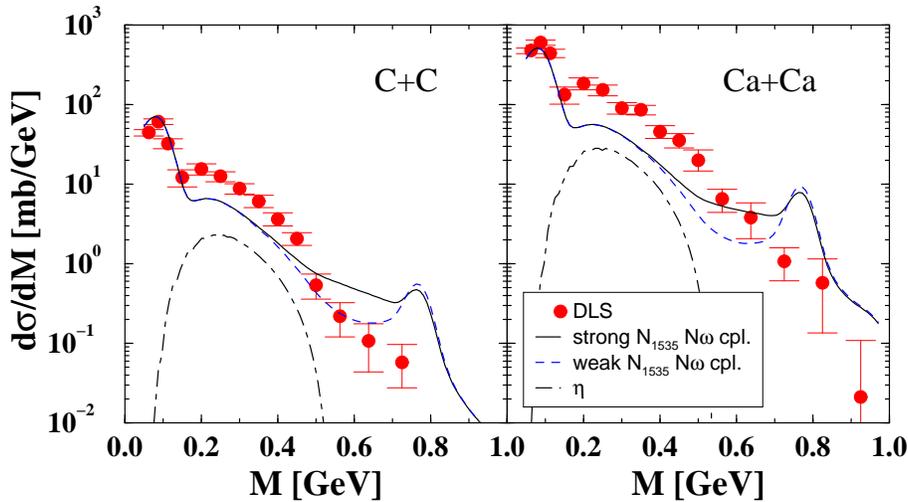,width=12.0cm}}}
\end{picture}
\end{minipage}
\vspace*{-1.5cm}
\hspace*{1.0cm}
\begin{minipage}[h]{40mm}
\caption{The dilepton spectrum in $C+C$ and $Ca+Ca$ reactions is 
compared to the DLS data \protect\cite{DLS}. The calculations are 
performed with a strong, respectively a weak $N^*(1535)-N\omega $ 
coupling.
}
\end{minipage}
\label{DLS_AA_fig1}
\vspace*{1.5cm}
\end{figure}
In the low mass region ($M=0.1\div 0.5$ GeV) we observe an underestimation 
of the DLS spectra  by a factor of $2\div3$. Thus in 
the present approach the underestimation of the DLS data 
is somewhat smaller than observed in the previous works of 
\cite{ernst} and \cite{BCRW98}. One reason for this is a larger 
$\eta$ contribution which is probably due to 
the iso-spin factor of 6.5 for the $np\rightarrow np\eta$ channel 
(compared to a factor of 2.5 used in \cite{BCRW98,BCM}). Other differences 
to the previous treatments \cite{ernst,BCRW98} are the following: 
In ref. \cite{BCRW98} the vector meson production 
was described by parameterizations of the $NN$ and $\pi N$ production channels 
while in the present approach these reactions run solely over the excitation of 
intermediate nuclear resonances. In \cite{ernst,BCRW98} only the 
$\Delta(1232)\rightarrow Ne^+ e^-$ Dalitz decay has explicitely been 
included. In addition, the decays of the nucleon resonances 
into vector mesons were treated till recently 
in the non-relativistic approximation \cite{BCM,pirner} 
and usually only one transition form factor was taken 
into account. From counting the independent helicity amplitudes it is clear that
a phenomenologically complete treatment requires
three transition form factors for spin $J \geq 3/2$ nucleon resonances and 
two transition form factors for spin-1/2 resonances.
Earlier attempts to derive a complete phenomenological 
expression for the dilepton decay of 
the $\Delta(1232)$ were not successful (for a discussion see \cite{krivo01}).
Despite of the details which differ in the various transport 
calculations (we included significantly more decay channels and apply an 
improved description of the baryonic resonance decays) the 
present results confirm qualitatively the underestimation of the 
DLS data at invariant masses below the $\rho/\omega$ peak \cite{ernst,BCRW98}.

A deviation to the results of \cite{ernst} and \cite{BCRW98} appears in the 
vicinity of the $\omega$ peak. Even after averaging over the experimental 
resolution the present results show a clear peak structure around 0.8 GeV 
which is absent in \cite{ernst,BCRW98}. However, in \cite{BCRW98} absorptive 
channels (e.g. $N\omega\rightarrow N\pi$ \cite{cassing99}) have been 
included which lead automatically to a collisional broadening of the 
in-medium vector meson width. Such a collision broadening is not 
included in the results shown in Fig. \ref{DLS_AA_fig1} but will 
separately be discussed in the next subsection. 
With respect to the  UrQMD calculations of \cite{ernst} our 
approach is in principle similar since vector mesons are produced through 
the excitation of nuclear resonances. However, in \cite{ernst}  the naive 
VMD was applied to treat the mesonic decays and the treatment is more qualitative, 
i.e. couplings were not particularly adjusted in order to describe 
$\rho$ and $\omega$ cross section as it was done in \cite{krivo02,omega02}. 
E.g. in  \cite{ernst} only the $N^*(1900)\rightarrow N\omega$ 
decay mode was taken into account which leads presumably to a significant 
underestimation of the $NN\rightarrow NN\omega$ cross section. 
\subsection{$\rho$- and $\omega$-meson in-medium widths}
In previous studies in-medium spectral functions of the $\rho$- and 
$\omega$-mesons were implemented into heavy-ion codes 
{\it ab initio} \cite{BCRW98}.  
At intermediate energies, the sensitivity of the dilepton spectra on 
the in-medium $\rho$-meson broadening is less pronounced as compared to
the $\omega$-meson. Estimates for the collision broadening 
of the $\rho$ in hadronic matter, i.e. dense nuclear matter or 
a hot pion gas, predict a collision width which is of the 
magnitude of the vacuum $\rho$ width. For the $\omega$, on the other hand, 
the vacuum width is only 8.4 MeV whereas in the medium it is 
expected to be more than one order of magnitude larger. However, 
the possibility of a strong in-medium modification of the $\omega$-meson 
has not attracted much attention in previous studies. The reason is 
probably due to the fact that the direct information on 
the $\omega$-meson channels from resonance decays, 
available through the multichannel $\pi N$ scattering analysis, 
is quite restricted. The present model provides an unified
description of the photo- and electro-production data and of the vector meson 
and dilepton decays of the nucleon resonances. 
It provides also a reasonable description of the vector meson 
and the dilepton production in elementary reactions ($p+p,p+d$) in 
the BEVALAC energy range. However, when applied to $A+A$ reactions 
the model leads to a very strong overestimation of the dilepton 
yield around the $\omega$-peak which suggests significant medium modifications 
of the $\omega$ contribution. At low energies, the vector 
meson production occurs due to decays of nucleon resonances.
The in-medium broadening of vector mesons can be understood within the 
framework of the resonance model. It has qualitatively two major 
consequences:
\begin{enumerate}
\item an increase of the nucleon resonance decay widths $R \rightarrow NV$ 

\item a decrease of the dilepton branchings $V \rightarrow e^+ e^-$  
due to the enhanced total vector meson widths. 
\end{enumerate}

These two effects are of opposite signs and can be 
completely described through 
appropriate modifications of the vector 
meson propagators entering into the $RN\gamma$ transition form 
factors $G_T(M^2)$. Within the eVMD framework it is sufficient to 
increase the total widths of the vector mesons. 
In a less formal way, the effect can be explained as follows: 
The differential branching 
\begin{equation}
dB(\mu,M)^{R\rightarrow NV} = 
\frac{d\Gamma_{\rm NV}^R (\mu,M)}{\Gamma_R (\mu)}
\label{bra}
\end{equation}
becomes usually larger with an increasing $V$ meson width which is 
due to the subthreshold character of the vector meson production through the 
light nucleon resonances. The dilepton branching of the nucleon resonances 
\begin{equation}
B(\mu)^{R\rightarrow Ne^+e^-} \sim
B(\mu)^{R\rightarrow NV} \frac{\Gamma_{V \rightarrow e^+e^-}}{\Gamma_V^{\rm tot}}
\label{bra1}
\end{equation}
is, on the other hand, inverse proportional to the total vector 
meson width $\Gamma_V^{\rm tot}$. Hence, 
an increase of the total width results in a decrease of the dilepton 
production rate. This effect is particularly strong for the 
$\omega$ since the in-medium $\omega$ width is expected to be
more than one order of magnitude greater than in the vacuum \cite{EBEK}. 
Although the estimates of ref. \cite{EBEK} were based on 
the standard VMD model which is contradictive with respect to the 
description of both, the $RNV$ and $RN\gamma$ branchings \cite{resdec,pirner,post01}, 
the qualitative conclusions concerning the magnitude of 
the in-medium $\omega$ broadening should be valid. 
A relatively large $\omega$ collision width is not too surprising. According to the 
$SU(3)$ symmetry the $\omega$ coupling to nucleons is 3 times greater than the $\rho$ 
coupling. One can therefore expect that 
at identical kinematical conditions the $N\omega$ cross section 
will be greater than the $N\rho$ cross section. Since the collision 
widths are proportional to the cross sections, the same 
conclusion holds for the collision widths.
The $\omega$ contribution is extremely sensitive to the reaction 
conditions in the course of the heavy ion collisions. While the 
increase of the total branching $B(\mu)^{R\rightarrow NV}$ depends 
on kinematical details one can expect that the suppression of the 
$\omega$ contribution due the enhanced total width $\Gamma_{\omega}^{\rm tot}$ 
is an one order of magnitude effect.
\begin{figure}[h]
\begin{minipage}[h]{105mm}
\unitlength1cm
\begin{picture}(10.,7.0)
\put(0.5,0){\makebox{\epsfig{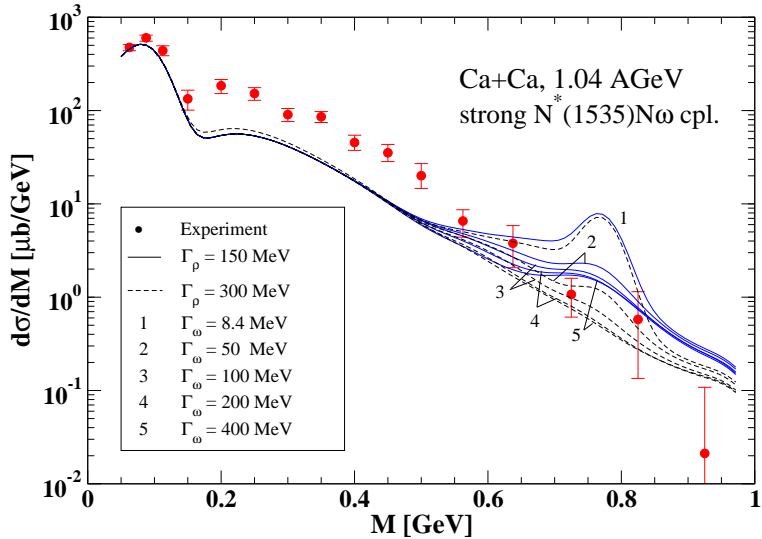}}}
\end{picture}
\end{minipage}
\vspace*{-1.5cm}
\hspace*{1.0cm}
\begin{minipage}[h]{60mm}
\caption{Dilepton spectra in $Ca + Ca$ collisions at 1.04 AGeV for different 
values of the in-medium $\rho$ and $\omega$ widths.
The solid curves correspond calculations where the $\rho$ width is 
kept at its vacuum value of $150$ MeV (no collision broadening). 
The dashed curves correspond to a total $\rho$ width of 300 MeV. 
In both cases the $\omega$ width is varied between  
$\Gamma_{\omega}^{tot} = 8.4\div 400$ MeV. 
}
\end{minipage}
\label{DLS_fig3}
\vspace*{1.5cm}
\end{figure}
The latter assumes 
an additional collision width of $\Gamma_{\rho}^{\rm coll} =150$ MeV 
which agrees with the estimates of refs. \cite{KKW,EIoffe,EBEK,KS}. 
In both cases the $\omega$ 
width is varied between $\Gamma_{\omega}^{\rm tot}$ = 8.4, 50, 100, 200, and 400 MeV. 
As already mentioned, the in-medium $\omega$ broadening is 
less studied. Thus we cover the possible range of in-medium 
values by the above parameter set.

First of all, it is important to realize that the region which is 
sensitive to in-medium modifications of the meson widths is distinct 
from the mass interval between $0.2\div 0.6$ GeV where the DLS puzzle is observed. 
This means that the problem to extract in-medium vector meson widths is 
isolated from the difficulties concerning the theoretical interpretation of the 
dilepton spectra below the $\rho/\omega$ peak. As expected, the dilepton spectra 
in the vicinity of the $\rho/\omega$ peak react very sensitive on modifications 
of the in-medium width. The reproduction of the DLS data requires an 
in-medium $\omega$ width which lies above 50 MeV for both, strong and weak 
couplings. As seen from Fig. 5, 
the best fits are obtained with $\Gamma_{\rho}^{\rm tot}$ = 300 MeV
and $\Gamma_{\omega}^{\rm tot}= 100\div 300$ MeV. With these values we 
reproduce in the strong $N^*(1535)N\omega$ 
coupling scenario the DLS data points around 
and 100 MeV below the $\rho/\omega$ peak within error bars. 
In the weak coupling scenario the DLS data are still slightly 
underestimated below the peak. However, the situation is not completely 
conclusive as discussed in detailk in \cite{shekther03}. 
Definite conclusions on the $N^*(1535)N\omega$ mode 
from dilepton yields in heavy ion reactions require more precise data 
which will be provided by HADES \cite{friese}. The present estimates 
can be interpreted as empirical values which 
are directly extracted from the experiment. The strength of the $\omega$ broadening
and the theoretical motivation through Eq. (\ref{bra1}) provide 
confidence for these estimates. 

If the average widths are fixed one can, on the other hand, extract 
an average cross section from the collision broadening condition 
$\Gamma^{\rm coll}_{VN} = \langle \rho_B\rangle  v\gamma \sigma_{VN}$. 
The average nuclear density at the vector meson production, respectively at the 
decay of the corresponding nuclear resonances $R$, is in minimal 
bias 1 AGeV $Ca+Ca$ reactions about 1.5 times the saturation density, i.e. 
$\langle \rho_B\rangle_{Ca+Ca} = 0.24~{\rm fm}^{-3}$ and slightly less for 
$C+C$ ($\langle \rho_B\rangle_{C+C} = 0.20~{\rm fm}^{-3}$). If one assumes 
now that the vector mesons are produced in an isotropic fireball with 
a temperature of $T\simeq 80$ MeV the extracted collisional width corresponds 
to an average  $\rho N$ cross section of about $\sigma_{\rho N}\simeq 30$ mb 
and $\sigma_{\omega N}\simeq 50$ mb for the $\omega$ 
($\Gamma^{\rm tot}_{\omega}=200$ MeV).
\subsection{Decoherence}
In refs. \cite{resdec,krivo02}, radially excited $\rho $- and 
$\omega $-mesons were introduced in the transition form factors $RN\gamma $ to
ensure the correct asymptotic behavior of the amplitudes in line with the
quark counting rules. Thereby we required a destructive interference
between the members of the vector meson families away from the poles of 
the propagators, i.e. the meson masses. In a dense medium the environment 
of the vector mesons can be regarded as a heat bath. Usually the 
different scattering channels of the interaction with a heat 
bath, i.e. the surrounding nucleons and pions, are summed up 
decoherently since the various channels acquire 
large uncorrelated relative phases. 
In such a case, the coherent 
contributions to the probability are random and cancel each other. 
We have in a sense macroscopically different intermediate states which do 
not interfere since small perturbations result in 
macroscopically large variations of the relative phases. 
The interaction of the vector mesons with the 
surrounding particles should therefore break up the 
coherence between the corresponding 
amplitudes for the dilepton production. The break up of the 
destructive interference results in an increase of the
total cross sections at low dilepton masses. In the following 
we want to investigate if the decoherence effect can explain 
the enhancement observed in the dilepton spectra at the BEVALAC 
experiment (DLS puzzle).

In the case of a full decoherence the vector meson contributions to the cross
section ${NN\rightarrow e^{+}e^{-}X}$ which run over nucleon resonances
must be summed up decoherently. This leads to the replacement 
\begin{equation}
|\sum_{k}{\cal M}_{Tk}^{(\pm )}|^{2}\rightarrow \sum_{k}|{\cal M}%
_{Tk}^{(\pm )}|^{2}~.  \label{E7}
\end{equation}
As a consequence, total decoherence will result in an 
enhancement of the resonance contributions due to the 
presence of the medium \cite{shekther03}. 

The decay probability for a resonance at 
distance $l_{C}$ in the interval $dl_{C}$ equals 
\begin{equation}
dW_{D}(l_{D})=e^{-l_{D}/L_{D}}\frac{dl_{D}}{L_{D}}~.
\end{equation}
The decay length for a resonance with lifetime $T_{D}$
equals $L_{D}=v\gamma T_{D}$, where $T_{D}=1/\Gamma $, $\Gamma $ being the
total vector meson vacuum width.
The collision probability at a distance $l_{C}$ in the interval $%
dl_{C}$ equals 
\begin{equation}
dW_{C}(l_{C})=e^{-l_{C}/L_{C}}\frac{dl_{C}}{L_{C}}~.
\end{equation}
The collision length $L_{C}$ is defined by the expression 
\begin{equation}
L_{C}=\frac{1}{\rho_B\sigma }  \label{coll}
\end{equation}
where $\sigma $ is the total $VN$ cross section and $\rho_B$ is the
nuclear density. The meson decay takes place before the first collision provided that 
$0<l_{D}<l_{C}$, so the probability of the coherent decay equals
\begin{eqnarray}
w =\int_{0}^{+\infty }\frac{dl_{C}}{L_{C}}e^{-l_{C}/L_{C}}%
\int_{0}^{l_{C}}\frac{dl_{D}}{L_{D}}e^{-l_{D}/L_{D}}  = \frac{L_{C}}{L_{C}+L_{D}}~.
\label{decprob1}
\end{eqnarray}
All mesons have in general different values
$L_{D}$ and $L_{C}$ and thus the coherent decay probabilities are
different as well. The collision broadening and the collision 
length are related through equations
\begin{equation}
e^{-l_C/L_C} = e^{- v t/L_C} = e^{-\Gamma^{\rm coll}_{V}t/\gamma }~~.
\label{reminder}
\end{equation}
Expression (\ref{reminder}) provides the the probability that 
a meson $V$ travels after its creation 
the length  $l_C$ through the medium without being scattered by 
the surrounding hadrons. In Eq.(\ref{reminder}), $v$ is velocity and $\gamma$
is the Lorentz factor. The collision length and width are thus related by
\begin{equation}
v /L_C = \Gamma^{coll}_{V}/\gamma ~~.
\label{reminder2}
\end{equation}
The collision length follows from the collisions widths 
which were extracted in the previous subsection. 
Since the collision widths are directly extracted from data, 
the $\rho$ and $\omega$ collision lengths which are necessary in order 
to determine the probabilities for a coherent dilepton emission 
can be obtained from (\ref{reminder2}). The estimates of the collision lengths 
for radially excited vector mesons are thereby assumed to be the same as 
for the ground-state vector mesons. The vacuum widths
of the radially excited mesons are larger than those of the ground 
state $\rho$ and $\omega$. As a consequence, the radially excited mesons 
show a tendency to decay coherently. The 
decoherence effect is most pronounced for the ground-state $\omega$-meson, 
since its vacuum width is particularly small. The $\omega$-meson decays 
in the medium almost fully decoherent, i.e. after its first collision 
with another hadron. This results in 
a modification of the $N^* \rightarrow Ne^+e^-$ decay rates of the 
$I=1/2$ resonances due to the destruction of the interference 
between the $I=0$ and $I=1$ transition form factors. Since for the 
considered reactions the matter is isospin symmetric, 
the break up of the $\rho-\omega$ coherence does not result 
in a significant change of the dilepton spectra. In this case 
the isoscalar-isovector interference terms cancel on average. 
The major effect arises from the break up of the interference 
between the $\omega$ and its radial excitations.

In \cite{shekther03} the influence of the decoherent 
summation of the intermediate 
mesonic states in the transition form factors was in detail
investigated. 
A totally decoherent summation of the mesonic amplitudes in the 
resonance decays enhances the dilepton yield generally by about a factor 
of two. In the low mass region this enhancement is able to match the 
DLS data. However, the scenario of a completely decoherent dilepton emission 
is rather unrealistic.
\begin{figure}[h]
\begin{minipage}[h]{125mm}
\unitlength1cm
\begin{picture}(12.,7.0)
\put(0.5,0){\makebox{\epsfig{file=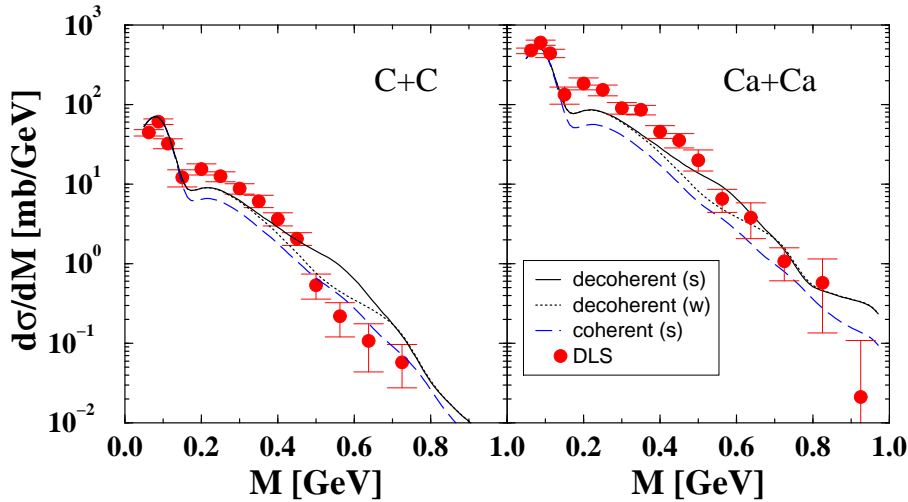,width=12.0cm}}}
\end{picture}
\end{minipage}
\vspace*{-1.5cm}
\hspace*{1.0cm}
\begin{minipage}[h]{40mm}
\caption{Influence of the microscopically determined decoherent dilepton emission 
in $C+C$ and $Ca+Ca$ reactions. A strong (s), respectively, weak (w) 
$N^*(1535)-N\omega $ coupling is used. For comparison also the 
coherent case (s) is shown. 
}
\end{minipage}
\label{DLS_AA_dec2}
\vspace*{1.5cm}
\end{figure}

In a realistic calculation shown in Fig. 6 the 
probabilities for coherent/decoherent dilepton 
emission are determined microscopically as outlined above, i.e. by 
the use of Eqs. (\ref{decprob1}-\ref{reminder2}). These 
use the 'optimal' values for the 
in-medium widths of $\Gamma_{\rho}^{\rm coll}=150, 
\Gamma_{\omega}^{\rm coll}=200$ MeV. The low mass dilepton yield 
is now enhanced by about 50\% by the decoherence effect which is, 
however, still too less to describe the DLS data. The interplay between 
the two in-medium effects, i.e. the collisional broadening and the 
decoherent dilepton emission is more complex. Decoherence leads also 
to an enhancement of the dilepton yield in the mass region between 
$0.4\div0.7$ GeV. Since the main decoherence effect occurs through 
the broken interference of the $\omega$ with its excited states, 
it is most pronounced in the dilepton contribution which stems from the 
$N^*$ resonance decays. This explains the difference between the 
two calculations assuming a strong/weak $N^*(1535)N\omega$ coupling 
in the mass range where possible off-shell $\omega$ contributions are 
now enhanced (strong coupling). However, definite conclusions on the 
strength of the $N^*(1535)N\omega$ coupling are still difficult to 
make at the present data situation. For the strong coupling the 
$Ca+Ca$ system is in agreement 
within error bars with the DLS data whereas in the lighter $C+C$ 
system the data are now overestimated and would favor the weak coupling. 
In both cases the agreement with the data is significantly improved in 
the low mass region. However, the considered decoherence 
effects are not completely sufficient in order to solve the DLS puzzle.  
The reason is that the microscopic determination of the decoherence 
probability favors the break up of the coherence between the $\omega$ 
and its excited states in the $N^*$ decays rather than the 
break up between the $\rho$ and its excited states in the $\Delta$ decays. 
The latter resonances are, however, those which contribute to most extent 
at low invariant masses. 
\section{Conclusion}
In the present work we provided a systematic description of vector 
meson and dilepton production in elementary $NN$ and $\pi N$ as well 
as in $A+A$ reactions. The reactions dynamics of the heavy ion collisions 
is described by the QMD transport model which was extended for the inclusion of 
nucleon resonances with masses up to 2 GeV. The vector meson production in 
elementary reactions is described through excitations of nuclear resonances 
within the framework of an extended VMD model. The model parameters were 
fixed utilizing electro- and photo-production data as well 
as $\pi N$ scattering analysis. Available data on the $\rho$ and $\omega$ 
production in $p+p$ and $\pi+N$ reactions are well reproduced. The same holds for 
the dilepton production in elementary $p+p$ and $p+d$ reactions. 

The situation becomes different turning to heavy ion collisions: 
In $C+C$ and $Ca+Ca$ reactions we observe in two distinct 
kinematical regions significant deviations from 
the dilepton yields measured by the DLS Collaboration. At small 
invariant masses the experimental data 
are strongly underestimated which confirms the observations made 
by other groups. Although accounting for the experimental resolution 
we observe further a clear structure of the $\rho/\omega$ peak which is  
not present in the data. Both features imply the investigation 
of further medium effects. 

The collisional broadening of the vector mesons suppresses the  
$\rho/\omega$ peak in the dilepton spectra. This allows to extract  
empirical values for the in-medium widths of the vector mesons. From  
the reproduction of the DLS data the following estimates for the 
collision widths $\Gamma_{\rho}^{\rm coll} = 150$ MeV and 
$\Gamma_{\omega}^{\rm coll} = 100 - 300$ MeV can be made. The in-medium values 
correspond to an average nuclear density of about 1.5 $\rho_0$. HADES will 
certainly help to constrain these values with higher precision. 

The second medium effect discussed here concerns the problem of 
quantum interference. Semi-classical transport models like QMD do generally 
not account for interference effects, i.e. they propagate probabilities 
rather than amplitudes and assume that relative phases cancel the 
interference on average. 
However, interference effects can play an important role for the dilepton 
production. In the present model the decay of nuclear resonances which is 
the dominant source for the dilepton yield, requires the destructive interference 
of intermediate $\rho$ and $\omega$ mesons with their excited states. 
The interference can at least partially be destroyed by the presence of 
the medium which leads to an enhancement of the corresponding dilepton 
yield. We proposed a scheme to treat the decoherence in the medium on 
a microscopic level. The account for decoherence improves the agreement with 
the DLS data in the low mass region. However, the magnitude of this effect is 
not sufficient to resolve the DLS puzzle completely. \\


\end{document}